\documentclass[floatfix,prl,aps,twocolumn,showpacs,preprintnumbers,amsmath,amssymb]{revtex4}

\usepackage{graphicx}
\usepackage{dcolumn}
\usepackage{bm}

\begin{document}

\title{Expansion of a Fermi gas interacting with a Bose-Einstein condensate}

\author{F. Ferlaino}
\author{E. de Mirand\'{e}s}
\author{G. Roati}
\author{G. Modugno}
\author{M. Inguscio}
\affiliation{LENS and Dipartimento di Fisica, Universit\`a di Firenze, and INFM\\
 Via Nello Carrara 1, 50019 Sesto Fiorentino, Italy }

\date{\today}

\begin{abstract}
We study the expansion of an atomic Fermi gas  interacting
attractively with a Bose-Einstein condensate. We find that the
interspecies interaction affects dramatically both the expansion
of the Fermi gas and the spatial distribution of the cloud in
trap. We observe indeed a slower evolution of the radial-to-axial
aspect ratio which reveals the importance of the mutual attraction
between the two samples during the first phase of the expansion.
For large atom numbers, we also observe a bimodal momentum
distribution of the Fermi gas, which reflects directly the
distribution of the mixture in trap. This effect allows us to
extract information on the dynamics of the system at the collapse.
\end{abstract}

 \maketitle

The dynamics of dilute quantum gases released from a trapping
potential, has provided crucial information on the interaction
properties of such systems. For instance, the anisotropic
expansion of a Bose-Einstein condensate (BEC) gave the first
direct evidence of the role of the boson-boson interaction and
provided an important test of the validity of the hydrodynamic
equations for a collisionless superfluid \cite{exp95,report}. Also
two-component Fermi gases at Feshbach resonances have been
recently found to exhibit a large anisotropic expansion
\cite{hoara,feshb} which can be described by hydrodynamic
equations \cite{menotti, pedri}. This has  confirmed the
achievement of a strongly interacting Fermi gas, which is one of
the requisites for fermionic superfluidity.
So far the experimental study of the expansion of quantum gases
has been focused essentially on  systems composed by  one- or
two-component gases of either bosonic or fermionic nature. Here we
extend the study to a Fermi-Bose system in the collisionless
regime. We use a two-species mixture composed by potassium and
rubidium atoms which exhibits a large interspecies attraction
($a_{BF}=$-410$^{+80}_{-80}$~$a_0$ \cite{Simoni}). We have already
found a modified expansion of the condensate in the mixture
\cite{roati}. We now study in detail the behavior of the Fermi gas
and we observe an anisotropic expansion in the presence of a BEC.
In particular, we find a slower evolution of the radial-to-axial
aspect ratio, which is mainly due  to a dynamical trapping
potential produced by the bosons during the early stage of the
expansion. We also observe an enhancement of the effect increasing
the overlap of the two clouds in trap and, for a large number of
condensed atoms, we find a bimodal momentum distribution  of the
expanded fermions which reflects directly their initial spatial
distribution in trap. Finally, we perform a study of the expansion
at the onset of the collapse of the Fermi gas \cite{collapse},
which is reached for even larger atom numbers in the mixture. This
provides information on the new equilibrium state reached by the
fermions after the collapse and on the timescale for the
equilibration.

\begin{figure}
\centerline{\includegraphics[width=8cm,clip=]{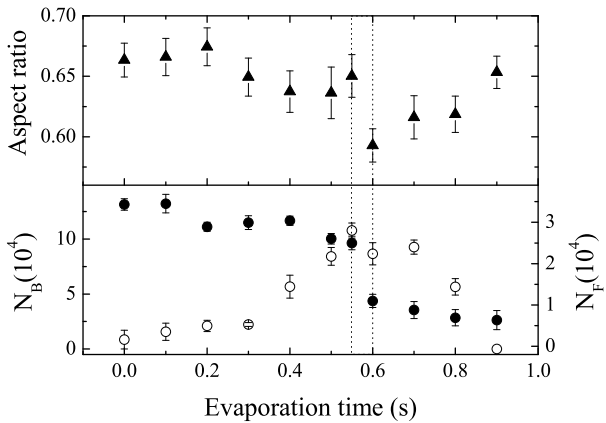}}
\caption{Evolution of the aspect ratio of the Fermi gas (solid
triangles) and of the number of bosons (empty circles) and
fermions (solid circles) at the collapse \cite{collapse}. During
the final stage of the bosons evaporation, the aspect ratio
decreases, and suddenly drops to a much lower value after the
collapse (highlighted region). The aspect ratio is measured at
$5\,$ms of expansion and the error bars are the standard deviation
of different measurements. } \label{Collasso}
\end{figure}

The procedure for creating an ultracold Fermi-Bose mixture of
$^{40}$K-$^{87}$Rb atoms has been described in detail in
Ref.~\cite{roati}. We cool typically a few $10^{4}$ fermions and
up to $10^{5}$ bosons to the quantum degenerate regime. Both
species are trapped in their stretched spin states, $|F=9/2,
m_F=9/2\rangle$ for K and $|2, 2\rangle$ for Rb. These states
experience the same trapping potential, with axial and radial
harmonic frequencies $\omega_{a}=2\pi \times 24\,$s$^{-1}$ and
$\omega_{r}=2\pi\times 317\,$s$^{-1}$ for K, while those for Rb
are a factor $(M_{Rb}/M_K)^{1/2}\approx 1.47$ smaller. For our
experimental parameters, the Fermi temperature for K is of the
order of $T_F=300\,$nK and the critical temperature for BEC of Rb
is $T_c=150\,$~nK. The experiments described here have been
performed at the lowest detectable temperature of the mixture,
corresponding to $0.2T_F$ for fermions and  $0.65T_c$ for bosons.
To study the expansion of the two species we suddenly switch off
the magnetic confinement and we detect the mixture at different
expansion times $t_{exp}$ by means of two-color absorption
imaging. From the imaging, we determine simultaneously the
radial-to axial aspect ratio of both samples.\\
\begin{figure}
\centerline{\includegraphics[width=7.5cm,clip=]{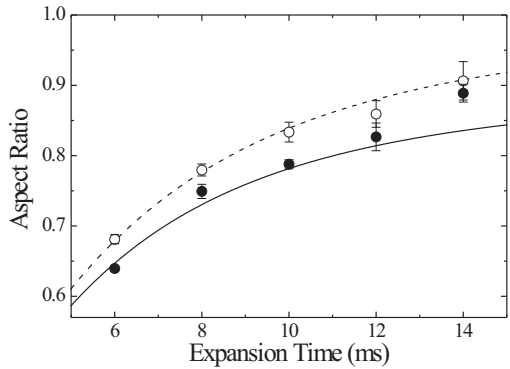}}
\caption{Modification of the expansion of a K Fermi gas due to the
interaction with a Rb BEC. The radial-to-axial aspect ratio of a
cloud of $4\times 10^4$ fermions  evolves more slowly in presence
of $3.5\times 10^4$ condensed bosons (solid circles) than in
presence of a similar number of uncondensed bosons (open circles).
Each data point is the average of five or six measurements. The
y-axis has been rescaled to take into account a distortion of our
imaging apparatus. The dashed line is the calculated expansion of
a pure Fermi gas, while the solid line is the prediction for an
interacting Fermi gas \cite{liu2}.} \label{arK}
\end{figure}
In a first experiment, we have studied the time evolution of the
aspect ratio of the Fermi gas interacting with a BEC. As shown in
Fig.~\ref{arK}, during the expansion the aspect ratio is always
smaller than that measured for a Fermi gas coexisting with a
dilute thermal cloud of bosons. The condensate, on the contrary,
inverts its aspect ratio more rapidly in presence of the Fermi
gas.
As we have already discussed \cite{roati}, the expansion of bosons
reveals an enhancement of density of the two gases in trap due to
the interspecies attraction, which results in a effective tighter
confinement. Indeed, if one assumes that also the interaction is
switched off at the release from the trap, this would lead to a
faster evolution of the aspect ratio. On the contrary, the
observed behavior of the Fermi gas gives now evidence that the
mutual attraction felt by the two species after the magnetic
confinement is switched off, plays a crucial role in the expansion
dynamics. Indeed, during the early stage of the expansion, each of
the two gases experiences a time-dependent trapping potential
produced by the other species. In this phase the negative
interaction energy is converted into kinetic energy which, in
general, can be unevenly distributed between the two samples.
Fig.~\ref{arK} shows that, in this system, the Fermi gas is taking
a large part of the interaction energy, which results in a largely
reduced kinetic energy in the radial direction, and therefore in
the slower evolution of the aspect ratio. This interpretation is
confirmed by a comparison with the theoretical study  of the
expansion of a Fermi-Bose mixture presented in Ref.~\cite{liu}. As
shown in Fig.~\ref{arK}, the observed evolution of the aspect
ratio of the Fermi gas is qualitatively well reproduced by the
theoretical prediction (solid line) calculated in the
collisionless regime. Note that this is actually the regime of our
system, for which we
measure a collisional rate smaller than the trap frequencies \cite{kdamping}.\\
We stress that our results reveal that the two species experience
differently the interspecies attraction due to their different
density distributions and energy scales. Indeed, the Fermi gas is
dominated by the attraction during the early phase of the
expansion while the evolution of the condensate is mostly affected
by the modification of the in-trap profile. This kind of
phenomenology is clearly not accessible in a two-component mixture
with a single density distribution. The numerical analysis of
Ref.~\cite{liu} and our experimental results suggest that the
interspecies interaction affects not only the expansion velocity
of the Fermi gas but also the asymptotic value of its aspect
ratio. To understand the general behavior, we derive an analytic
expression of the asymptotic value of the aspect ratio using a
scaling approach \cite{spazio}.
Following Ref.~\cite{liu} and to first order in
$\epsilon=\omega_a/\omega_r$, we assume that the radii of the
bosonic cloud evolve in the radial direction according to
$R_r^b(t)=R_r^b(0)\sqrt{1+(\beta \omega^b_r)^2t^2}$, where $\beta$
is an effective renormalization of the bosons trap frequency
$\omega^b_r$, and in the axial direction like $R_a^b(t)\approx
R_a^b(0)$. For our cigar shaped harmonic trap, we find that the
aspect ratio approaches an asymptotic value
\begin{equation}
\frac{R_r}{R_a}\rightarrow\frac{
\alpha}{\sqrt{1-\frac{3}{2}\chi_a}},
\label{Asym}
\end{equation}
where $\omega'_r=\alpha\omega_r$ is the effective trap frequency
of the fermions in the radial direction, with
$\alpha^2=(\beta^4/3\lambda^4\chi_r)(1-\sqrt{1-(6\chi_r-9\chi^2_r)\lambda^4/\beta^4})
$ and $\lambda=\omega_r/\omega^b_r$. The coupling function
$\chi_{i=\{a,r\}}$ characterizes the interspecies interaction:
\begin{equation}
\chi_i=\frac{E_{int}}{E_{ho}}+\frac{g_{bf}}{E_{ho}}\int d^{3}x
\frac{\partial n_f}{\partial x_i}x_i n_b,
\end{equation}
with $E_{int}=g_{fb}\int d^3xn_fn_b$, $E_{ho}=\int d^3x
V_{ho}n_f$, and $V_{ho}$ the harmonic confinement.
Eq.~(\ref{Asym}) shows that the  expansion of the fermions is
directly coupled to the one of bosons via the  parameter $\beta$.
The aspect ratio reaches a value which depends only on the
interaction with the condensate in trap and on $\beta$. Note that
Eq.~(\ref{Asym}) reduces to the one found in Ref.~\cite{menotti}
for two coupled clouds with the same density distribution
($n_b=n_f$). In this case, the coupling function takes the simpler
form $E_{int}/E_{hoi}$ and, for an attractive interaction, the
aspect ratio of the two clouds approaches always a value smaller
than one. In a Fermi-Bose mixture, this result is no longer true
and the asymptotic value for the Fermi gas can be smaller or
larger than one, depending on the bosons expansion and on the
coupling function $\chi$. For our specific parameters, we find
that $R_r/R_a\rightarrow 0.8$, which is smaller than one, in
accordance with our experimental results \cite{Michele}. Note that
Eq.~(\ref{Asym}) indicates that in general it is not possible to
extract the sign of the interspecies interaction just from the
asymptotic value of the aspect ratio.
\\
\begin{figure}
\centerline{\includegraphics[width=6.8cm,clip=]{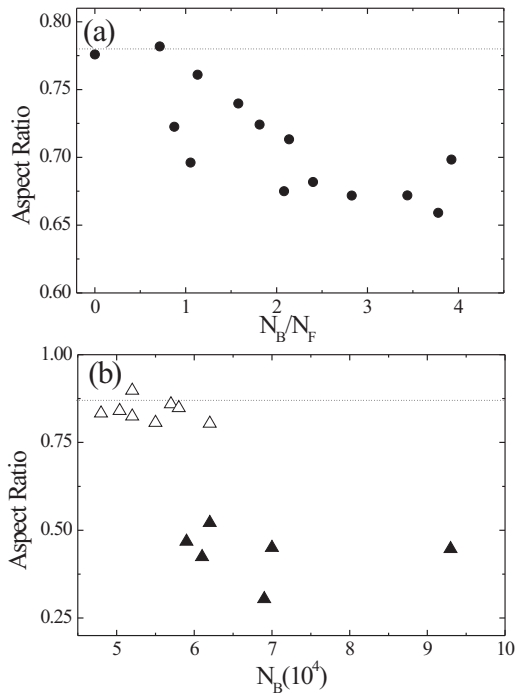}}
\caption{Aspect ratio of the Fermi gas at $8\,$ms of expansion
varying the atom numbers in the mixture. The interspecies overlap
region is enhanced either (a) decreasing $N_F$, for $N_B\simeq
6\times 10^4$ or (b) increasing $N_B$ with $N_F\simeq 2.4\times
10^4$. Above $N_B=6\times 10^4$, we observe the appearance of a
bimodal radial (see Fig.~\ref{densita}(b)=) . The aspect ratio of
the central component (solid triangles) is obtained fitting both
the radial and axial distribution with a two-peak gaussian
function. The dashed lines indicate the nominal value of the
aspect ratio for a pure Fermi gas. } \label{arN}
\end{figure}
\begin{figure}
\centerline{\includegraphics[width=6cm,clip=]{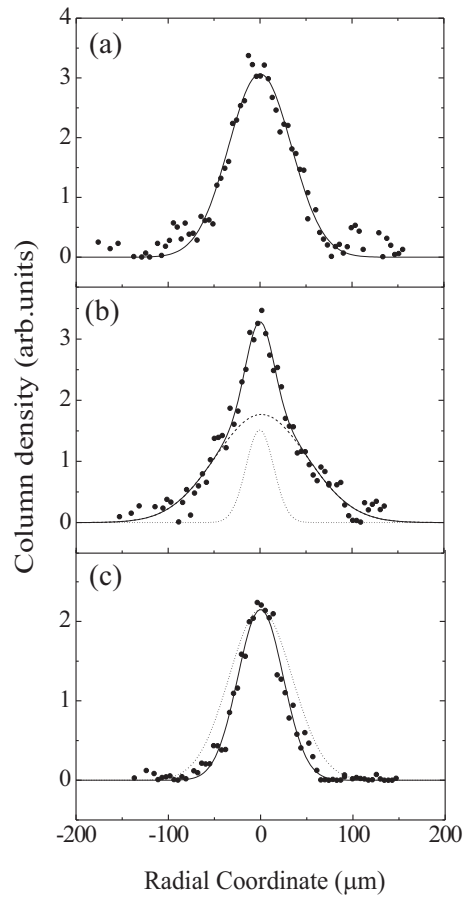}}
\caption{Radial momentum distribution of the Fermi gas, detected
after $12\,$ms of expansion, for different atom number in the
mixture. For $N_F=2\times 10^4$, (a) we observe that below
$N_B=6\times 10^4$  the distribution is slightly affected by
interspecies potential, while (b) above $N_B=6\times 10^4$ a
bimodal structure appears. (c) After the collapse, the remaining
$1.2\times 10^4$ fermions coexisting with  $7.4\times 10^4$ bosons
exhibit a narrower distribution than the non-interacting gas
(dotted line).} \label{densita}
\end{figure}
We have investigated how the expansion of the Fermi gas depends on
the atom numbers in the mixture. We find that the aspect ratio
decreases reducing  the number of fermions at a constant number of
bosons, as shown in Fig.~\ref{arN}(a). This behavior is not
surprising because a decrease on $N_F$ corresponds on the one hand
to a reduction of the overall density and therefore of the
interaction energy. On the other hand it leads to a decrease of
the Fermi radius ($R_F\propto(N_F)^{1/6}$), resulting in a better
spatial overlap of the two samples. An alternative way to increase
the extension of the overlap region is to increase $N_B$ at
constant $N_F$ which also leads to a larger density in the overlap
region. In this case, we observe the appearance of a double
distribution in the radial profile of fermions with a narrow peak
surrounded by a broader distribution, as illustrated in
Fig.~\ref{densita}. This double distribution gives the first
direct evidence of the enhancement of the fermions density in trap
arising from the attraction with bosons. We attribute indeed the
narrow distribution to the fermions trapped into the condensate
while the broader distribution is occupied by the more energetic
atoms outside the overlap volume. Only the central component
experiences the trapping potential produced by the bosons during
the first phase of the expansion and its expansion is slowed down
accordingly. The other component instead expands normally. The
momentum distribution therefore reflects the spatial distribution
in the trap. Note that, in absence of an interspecies attraction
during the expansion, the  effect in trap would vanish in a
characteristic time of the order of $1/\omega_r$ and one could
detect the enhancement of density only studying the spatial
distribution in trap, which is not experimentally accessible.  We
observe the bimodal distribution only in the radial direction,
confirming that the interaction energy between the two clouds is
exchanged mainly in the more tightly confined direction
\cite{exp95,hoara}.\\ In Fig.~\ref{arN}(b), we show a quantitative
study of the aspect ratio of fermions in the regime of large boson
number. For low $N_B$, where we are not able to distinguish the
two distributions, the aspect ratio is only moderately  affected
by the Bose-Fermi interaction. For sufficiently large $N_B$, the
two different distributions become visible and we can measure the
aspect ratio of just the fraction of fermions interacting with the
bosons. In this case, we detect a reduction of the aspect ratio down to one half of the non-interacting value. \\
Finally, by further increasing the atom numbers in the mixture, we
study the aspect ratio of the Fermi gas at the occurrence of the
collapse. We recall that an instability of the Fermi gas appears
in this system when the interspecies attraction is no longer
balanced by the Pauli repulsion \cite{collapse}. In our system,
the collapse is detected as a sudden drop of $N_F$ to typically
less than half its original value. We now observe that, after the
collapse, the remaining fermions exhibit a radial distribution
usually narrower than the non-interacting one, as shown in
Fig.~\ref{densita}(c). Actually, the aspect ratio of the fermions
in this condition is up to $30\%$ lower than the one expected for
a pure Fermi gas and of the same order of the one measured for a
stable mixture with comparable atom numbers. This suggests that
after the collapse the system has reached a new equilibrium
distribution in which most of the fermions are immersed in the
condensate feeling a large interspecies interaction during the
expansion from the trap. Our observation is somehow surprising,
since one could expect that during the collapse exactly the
fermions in the high density region within the condensate are lost
through inelastic processes while the remaining fermions are
mostly located in the outer region. This would correspond either
to a faster expansion with a larger aspect ratio, or to an
oscillating Fermi gas strongly out of the equilibrium which we do
not observe in the experiment. To investigate the timescale of the
equilibration process of the Fermi gas, we have analyzed the
evolution of the aspect ratio of the fermions across the collapse
for the measurements already reported in Ref.~\cite{collapse}. As
shown in Fig.~\ref{Collasso}, increasing $N_B$ we first observe a
small decrease of the aspect ratio followed by a jump to a lower
value just in correspondence to the collapse. The aspect ratio
then slowly tends to the unperturbed value as the BEC is
completely evaporated. Since we invariably see a smaller aspect
ratio just after the collapse, we can conclude that the system
finds a new equilibrium distribution in a time scale  smaller than
$50\,$ms, which is our effective time resolution in the
experiment.

In conclusion, we have studied the effect of the attractive
interaction with a BEC on the expansion of a Fermi gas. We observe
an anisotropic expansion of the cloud which consists in a slower
evolution of the aspect ratio which approaches an asymptotic value
smaller than the unity. The opposite behavior that we find for the
BEC and, most of all, the observation of a bimodal momentum
distribution of the Fermi gas confirm that the phenomenology of a
system composed by two species with different spatial distribution
is much richer of than that expected for a one-distribution gas
\cite{menotti}. Here we also show that the peculiar expansion of
the Fermi gas can be used to extract information on the atoms
which are overlapped with the BEC in trap. In future experiment it
would be interesting to study the expansion of the mixture at the
crossover from the collisionless to the hydrodynamics regime by
tuning the interaction via Feshbach resonances. These study on
strongly interacting Fermi-Bose mixture can be relevant also for
the recently observed mixtures of fermionic atoms and condensed
bosonic molecules \cite{becMolecolare}.\\
We are grateful to X.~J.~Liu, H.~Hu, M.~Modugno and H.~ Ott for
fruitful discussions and to F.~Riboli for contributions to the
experiment. This work was supported by MIUR, by EU under the
Contract HPRICT1999-00111, and by INFM, PRA ``Photonmatter''.


\begin{thebibliography}{99}
\bibitem{exp95} M.~H.~Anderson {\it et al.}, Science {\bf 269},
198 (1995); K.~B.~Davis {\it et al.} , Phys. Rev. Lett. {\bf 75}
3969 (1995).

\bibitem{report} F.~Dalfovo, S.~Giorgini, L.~P.~Pitaevskii, S.~Stringari, Rev. Mod. Phys. {\bf 71}, 463 (1999).

\bibitem{hoara} K.~M.~O'Hara {\it et al.}, Science {\bf 298}, 2179 (2002).

\bibitem{feshb} C.~A.~Regal, and D.~S.~Jin, Phys. Rev. Lett. {\bf 90}, 230404
(2003); T.~Bourdel {\it et al.}, Phys. Rev. Lett. {\bf 91}, 020402
(2003).


\bibitem{menotti} C.~Menotti, P.~Pedri, and S.~Stringari, Phys. Rev. Lett. {\bf 89}, 250402
(2002).

\bibitem{pedri} P.~Pedri, D.~Guéry-Odelin, and S.~Stringari
Phys. Rev. A {\bf 68}, 043608 (2003)
\bibitem{Simoni} A.~ Simoni, F.~ Ferlaino, G.~ Roati, G.~ Modugno, and M. Inguscio
Phys. Rev. Lett. {\bf 90}, 163202 (2003).

\bibitem{roati} G.~Roati, F.~Riboli, G.~Modugno,
and M.~Inguscio, Phys. Rev. Lett. {\bf 89}, 150403 (2002).

\bibitem{collapse} G.~Modugno {\it et al.}, Science {\bf 297}, 2200 (2002).

\bibitem{liu} H.~Hu, X.~J.~Liu, and M.~Modugno Phys. Rev. A {\bf 67}, 063614
(2003).
\bibitem{kdamping}  F.~Ferlaino {\it et al.}, J. Opt. B {\bf 5}, s3 (2003).

\bibitem{spazio} Details of our calculation will be given in a longer
report.

\bibitem {liu2}The theoretical curve has been obtained by a numerical
simulation provided by X.J. Liu for our experimentals parameters.
The numerical calculation are performed with $a_{BF}=-330 a_0$
instead of $a_{BF}=-410 a_0$, which is the value that better fits
also the expansion of bosons, as discussed in Ref.~\cite{liu}.

\bibitem{Michele} M.~Modugno private comunication (2003).

\bibitem{becMolecolare} S.~Jochim {\it et al.}, Science Express, (10.1126/science.1093280)(2003);
M.~Greiner, C.~A.~Regal, and D.~S.~Jin , Nature,
(doi:10.1038/nature02199) (2003); M.~W.~Zwierlein {\it et al.},
cond-mat/0311617 (2003).
\end{thebibliography}
\end{document}